\documentclass[final,1p,times]{elsarticle}

\usepackage{amssymb}
\usepackage{amsmath}

\usepackage[ruled,linesnumbered]{algorithm2e}

 \newtheorem{theorem}{Theorem}

\begin{document}

\begin{frontmatter}


\title{Balanced Low-Complexity and Flexible Error-Correction List Flip Decoding for Polar Codes\tnoteref{label0}}


\author[label1]{Yansong Lv} 
\ead{lys_communication@cuc.edu.cn}

\author[label2]{Jingxin Dai}
\ead{daijingxin@cuc.edu.cn}

\author[label2]{Yuhuan Wang\corref{cor1}}
\ead{wangyuhuan@cuc.edu.cn}
\cortext[cor1]{Corresponding author: Yuhuan Wang}

\author[label2]{Hang Yin}
\ead{yinhang@cuc.edu.cn}

\author[label2]{Zhanxin Yang} 
\ead{yangzx@cuc.edu.cn}

\affiliation[label1]{organization={Information and Communication Engineering Institute, Communication University of China},
            postcode={100024},
            state={Beijing},
            country={China}}

\affiliation[label2]{organization={State Key Laboratory of Media Convergence Communication, Communication University of China},
            postcode={100024},
            state={Beijing},
            country={China}}


\begin{abstract}
Benefiting from performance advantages under short code lengths, polar codes are well-suited for certain scenarios, such as the future Internet of Things (IoT) applications that require high reliability and low power.
Existing list flip decoders can efficiently further enhance the error-correction performance of polar codes with finite code lengths, particularly the dynamic successive cancellation list flip (D-SCLF) decoder with flexible high-order error-correction capability (FHECC).
However, to the best of our knowledge, current list flip decoders cannot effectively balance complexity and error-correction efficiency.
To address this, we propose a parity-check-aided D-SCLF (PC-DSCLF) decoder.
This decoder, based on FHECC and the characteristics of the list flip decoding process, introduces a simplified flip metric and a hybrid check scheme, along with a decoding method that supports the check scheme, enabling it to retain FHECC while achieving low complexity.
Simulation results show that the proposed PC-DSCLF decoder achieves up to a 51.1\% average complexity reduction compared to the D-SCLF algorithm with distributed CRC for $PC(512, 256+24)$.
\end{abstract}



\begin{keyword}

polar code \sep low complexity \sep list flip decoder \sep parity check \sep simplified flip metric




\end{keyword}

\end{frontmatter}



\section{Introduction}
\label{s1}
Polar codes \cite{[1]}, a channel coding technique with significant performance advantages under short code lengths, have been adopted in the channel coding scheme for control information in the 5G enhanced Mobile Broadband (eMBB) scenario \cite{[102]}, and exhibit strong competitiveness in scenarios requiring high reliability and low power, such as future Internet of Things (IoT) applications \cite{[103],[104]}.
For instance, to satisfy the ultra-low power consumption requirements of devices in Narrowband IoT (NB-IoT) and enhanced Machine Type Communication (eMTC) \cite{[105]}, various enhanced repeated polar codes have been successively proposed to ensure reliable communication under extremely low signal-to-noise ratio (SNR) conditions \cite{[106],[107],[108]}.
Additionally, to address the diverse data reliability requirements of heterogeneous IoT devices, researchers have developed polar code algorithms with unequal error protection (UEP) characteristics \cite{[109],[110]}.
Moreover, to satisfy the demands for low-latency and high-reliability communication between IoT devices \cite{[111],[112]}, researchers have proposed various improved polar code algorithms from multiple perspectives, including decoding strategies \cite{[113],[114]}, encoding techniques \cite{[115]}, and joint optimization methods \cite{[116]}.

Enhancing the performance of decoders is undoubtedly the key to ensuring the competitiveness of polar codes. 
Polar codes with the successive cancellation (SC) decoder are the first channel codes that can provably achieve channel capacity with infinite code length \cite{[1]}. 
To further improve error-correction performance under finite code length, several enhanced SC-based decoders have been proposed, including the SC list (SCL) decoder \cite{[3], [4]} and the cyclic redundancy check (CRC)-aided SCL (CA-SCL) decoder \cite{[5], [6]}. 
These two advanced decoders utilize a list of multiple candidate paths to replace the single candidate path in the SC decoder, thereby significantly improving error-correction performance. 
In particular, the CA-SCL decoder achieves superior error-correction performance compared to the SCL decoder by incorporating a CRC.
However, the use of multiple candidate paths increases computational complexity (and energy consumption) as well as storage requirements. 
To address these challenges, researchers have proposed various improved SCL-based decoders, including the adaptive SCL decoder \cite{[7]}, log-likelihood ratio (LLR)-based SCL decoder \cite{[8]}, fast simplified SCL decoder \cite{[9]}, and the list-pruning decoder \cite{[10]}. 
Nevertheless, the improvements in these SCL-based decoders have limited contributions to error-correction performance.

To further enhance the error-correction performance of these SCL-based decoders, various list flip decoders have been proposed.
The SCL bit-flip (SCL-BF) decoder \cite{[11]}, the first list flip algorithm for polar codes, introduces additional attempts using bit-flip operations \cite{[12]} to recover the correct path discarded by a failed list decoder, thereby achieving better error-correction performance than the CA-SCL decoder with the same list size.
To improve the efficiency of bit-flip operations, \cite{[13]} and \cite{[14]} successively proposed the SCL decoder with shift-pruning (SCL-SP) and the SCL-Flip decoder. 
However, these list flip decoders lack flexible high-order error-correction capability (FHECC), as they can only execute high-order error correction after completing all low-order error correction attempts. 

To address this issue, \cite{[20]} introduced a novel flip metric and developed a new list flip decoder with FHECC, referred to as the dynamic successive cancellation list flip (D-SCLF) decoder. 
Unlike earlier list flip decoders, the D-SCLF decoder estimates high-order error probabilities using the newly designed flip metric and directly performs high-order error-correction. 
This innovation significantly enhances the efficiency of list flip operations and improves the error-correction performance of list flip decoders.
However, similar to other list flip decoders, the D-SCLF decoder suffers from excessive computational complexity in high SNR environments, which limits its practical applicability. 
Although several recently developed list flip decoders, such as deep-learning-based decoders \cite{[16],[18],[181]} and latency-reducing decoders \cite{[15],[17]}, provide partial improvements, they consistently fail to achieve a balance between FHECC and computational complexity in low SNR conditions.
Therefore, designing a list flip decoder that simultaneously achieves low average computational complexity and maintains FHECC represents a highly practical solution for meeting the demands of high-reliability and low-energy communication.

In this work, we propose a low complexity list flip decoder with FHECC, called the parity-check-aided D-SCLF (PC-DSCLF) decoder. The main contributions of our work are as follows:
\begin{itemize}
\item 
Firstly, we introduce a simplified flip metric that preserves FHECC.
Compared to the latest simplified flip metric in \cite{[181]}, the proposed one reduces nonlinear operations in the D-SCLF decoder while maintaining its FHECC.

\item 
Secondly, we present a hybrid-check-based decoding scheme.
To ensure efficient error detection and early termination, we established two requirements and developed a specific hybrid-check scheme based on them.
Building on this, we propose a hybrid-check-aided decoding scheme.
Simulation results demonstrate that the D-SCLF decoder with the proposed hybrid-check-based decoding scheme achieves up to a 64.1\% average complexity reduction compared to the D-SCLF algorithm with undistributed CRC bits, for $PC(512, 256+24)$.

\item 
Finally, we propose the PC-DSCLF decoder by combining the proposed simplified flip metric and the hybrid-check-based decoding scheme. 
Simulation results show that the proposed PC-DSCLF decoder achieves up to a 51.1\% average complexity reduction compared to the D-SCLF algorithm with distributed CRC, for $PC (512, 256+24)$.
\end{itemize}

The remainder of the paper is organized as follows. Section II briefly overviews the encoding method and main SCL-based decoders. Section III describes and analyzes the details of the proposed decoders. In Section IV, the simulation results are illustrated and discussed. Finally, some conclusions are highlighted.

\section{Preliminaries}
\label{s2}

In this section, we provide a brief overview of the encoding method and the main SCL-based decoders for a polar code $PC (N, K+n_{\mathbf{c}})$, where the code length is $N$, the CRC length is $n_{\mathbf{c}}$, the information bits length is $K$, and the code rate is $R = K/N$. 
It is important to note that the decoders introduced in this section only utilize CRC, meaning their total number of check bits is equal to $n_{\mathbf{c}}$.

Moreover, in this paper, lowercase and uppercase letters denote scalars (e.g., $y$ and $Y$), bold lowercase letters denote vectors (e.g., $\mathbf{y}$), bold uppercase letters denote matrices (e.g., $\mathbf{Y}$), script uppercase letters represent sets (e.g., $\mathcal{Y}$),
$\mathcal{Y}[i]$ denotes the $i^{\text{th}}$ element of the set $\mathcal{Y}$, $\mathbf{y}[i]$ represents the $i^{\text{th}}$ element of the vector $\mathbf{y}$, and $y_1^N = \{ y_1, y_2, \ldots, y_N \}$ denotes a specific vector.

\subsection{Polar Encoding Method}

The polar encoding method \cite{[1]} is defined as
\begin{equation}
	x_{1}^{N}=u_{1}^{N}\mathbf{B}_{N}\mathbf{F}^{\otimes n},
\end{equation}
where $x_1^N$ denotes the encoded vector, $\mathbf{B}_N$ is a bit-reversal permutation matrix, $\otimes$ denotes the Kronecker product, and $\mathbf{F}=[\begin{smallmatrix}1 & 0 \\ 1 & 1 \end{smallmatrix}]$.
Moreover, $u_1^N$ is the encoding vector, which consists of two subsets: 
    \begin{itemize}
		\item $\mathbf{u}_{\mathcal{A}^c}$: the frozen bit sequence, where $\mathcal{A}^c$ records the indices of the frozen bits in $u_1^N$ in ascending order.
		\item $\mathbf{u}_\mathcal{A}$: the non-frozen bit sequence, where $\mathcal{A}$ records the indices of the non-frozen bits in $u_1^N$ in ascending order.
    \end{itemize}

\subsection{CA-SCL Decoder}

The CA-SCL decoder \cite{[4],[5]} is a well-known decoder for polar codes due to its superior performance. 
The primary reason for its excellent performance is the aid of CRC in selecting the best decoding path.

Assume $L$ is the total number of candidate paths. 
In CA-SCL decoding, the path metric (PM) is used to measure the reliability of these paths, and can be computed as \cite{[8]}:
\begin{equation}
		{PM}_{l}^{(i)}=  \left\{\begin{matrix} {PM}_{l}^{(i-1)}, \text{if}\quad \hat{u}_{i}[l]=\delta (\lambda_{N}^{(i)}[l]),                                        \\
               {PM}_{l}^{(i-1)}+\left| \lambda_{N}^{(i)}[l] \right|,  \text{otherwise}.  \\
		\end{matrix}\right.
\end{equation}
where $\delta(x)=\frac{1}{2}(1-sign(x))$, and $\lambda_{N}^{(i)}[l]$ refers to the log-likelihood ratio (LLR) value of the $i^{\text{th}}$ bit in the $l^{\text{th}}$ candidate path. 
${PM}_l^{(i)}$ represents the PM value of the $i^{th}$ bit in the $l^{th}$ candidate path, and ${PM}_l^{(0)}=0$. 
A candidate path with a larger PM is more likely to be an incorrectly estimated path. 

When decoding the $i^{\text{th}}$ bit and $i\in\mathcal{A}'$, CA-SCL decoder utilizes a list $\mathcal{L}_{best}^{(i)}$ to reserve all candidate paths, where $\mathcal{A}'$ refers to the set consisting of the first $\log_2(L)$ non-frozen bits. 

When decoding the $i^{\text{th}}$ bit and $i\in\mathcal{A}\setminus\mathcal{A}'$, the $L$ paths in  $\mathcal{L}_{best}^{(i-1)}$ are expanded to $2L$ sub-paths, forming an expanded list $\mathcal{L}^{(i)}$. 
Then, $\mathcal{L}_{best}^{(i)}$ can be achieved by selecting $L$ paths with smaller PM from  $\mathcal{L}^{(i)}$. 

After all bits are decoded, the CA-SCL decoder performs CRCs on paths in the  $\mathcal{L}_{best}^{(N)}$. 
If all CRC fails, the CA-SCL decoder outputs the candidate path with the smallest PM in $\mathcal{L}_{best}^{(N)}$. 
Otherwise, it outputs the candidate path with the smallest PM among those passing CRCs. 

\subsection{List Flip Decoder}

To improve the error-correction performance of the CA-SCL decoder, an SCL bit-flip (SCL-BF) decoder \cite{[11]}, the first list-based flip decoder, was proposed.  
This decoder identifies positions prone to errors in CA-SCL decoding attempts by constructing a revised critical set and corrects these identified errors through additional decoding attempts. 
Simulation results show that, the BLER performance of the SCL-BF decoder outperforms the CA-SCL decoder by approximately 0.12-0.25 dB in the medium and high SNR regions, with negligible additional complexity.

To further improve the accuracy of identifying error bits, \cite{[14]} proposed the SCL flip (SCL-Flip) decoder with a proposed novel flip metric.  
The new metric can be computed as:
\begin{equation}
	E_{\alpha}^{(i)} = \ln\frac{\sum_{l=1}^{L}e^{-PM_l^{(i)}}}{{\left(\sum_{l=1}^{L}e^{-PM_{l+L}^{(i)}}\right)}^\alpha},
	\label{eq:Ei} 
\end{equation}
where $E_{\alpha}^{(i)}$ denotes the flip metric value of the $i^{\text{th}}$ bit, and $\alpha$ is a coefficient used to compensate for biased estimation caused by error propagation.  
Specifically, when $\alpha=1$, Eq. \eqref{eq:Ei} can be simplified to $E_{1}^{(i)}$:
\begin{equation}
	E_{1}^{(i)} = \ln\frac{\sum_{l=1}^{L}e^{-PM_l^{(i)}}}{\sum_{l=1}^{L}e^{-PM_{l+L}^{(i)}}},
	\label{eq:E1}
\end{equation}
A lower $E$ value indicates a higher likelihood of errors occurring.  
Therefore, the SCL-Flip decoder prioritizes flipping information bits with lower $E$ values.  
However, a lower $E$ value does not necessarily indicate a higher likelihood of the first error occurring.

To address the problem of accurately estimating the position of the first error in list decoders, \cite{[20]} proposed a D-SCLF decoder.  
The core idea of the D-SCLF decoder is represented by the following equations:
\begin{equation}
	\varepsilon^{(i)} = \{u_1^{i-1} \in \mathcal{L}_{best}^{(i-1)}, u_1^{i} \notin \mathcal{L}_{best}^{(i)} \},
\end{equation}
and
\begin{equation}
	P(\varepsilon^{(i)}|y_1^N) = P_e^{(i)} \cdot \prod_{k < i, k \in \{\mathcal{A} \setminus \mathcal{A}'\}} (1 - P_e^{(k)}),
\end{equation}
where $\varepsilon^{(i)}$ represents the event that the first error occurs in the $i^{\text{th}}$ bit, $P(\varepsilon^{(i)}|y_1^N)$ represents the probability of this event, and $P_e^{(i)} = \frac{1}{1 + e^{\beta \cdot E_{1}^{(i)}}}$.  

In particular, the paper also estimates the probability of higher-order errors, expressed as:
\begin{equation}
	P(\varepsilon^{(i)}|y_1^N, \mathcal{S}_t) = P_e^{(i)} \prod_{k \in \mathcal{S}_t} P_e^{(k)} \prod_{k < i, k \in \{\mathcal{A} \setminus \mathcal{A}'\} \setminus \mathcal{S}_t} (1 - P_e^{(k)})
\label{eq:PeSt}
\end{equation}
where $\mathcal{S}_t$ records all flip indices for the $t^{\text{th}}$ additional decoding attempt, and $P(\varepsilon^{(i)}|y_1^N, \mathcal{S}_t)$ represents the probability of the $i^{\text{th}}$ bit being erroneous after list flips have been executed at the positions recorded in $\mathcal{S}_t$.  
Furthermore, this probability can be transformed into the following expression:
\begin{equation}
	\begin{aligned}
		M_{\beta}^{(i)} &= -\frac{1}{\beta} \ln(P(\varepsilon^{(i)}|y, \mathcal{S}_t)) \\
		&= E_{1}^{(i)} + \sum_{k \in \mathcal{S}_t} E_{1}^{(k)} + \sum_{k \leq i, k \in \mathcal{A} \setminus \mathcal{A}'} f_{\beta}(E_{1}^{(k)}),
		\label{eq:LnPeSt}
	\end{aligned}
\end{equation}
where $f_{\beta}(x) = \frac{1}{\beta} \ln(1 + e^{-\beta x})$, and $\beta$ is a compensation coefficient similar in function to $\alpha$. 

Based on Eq. \eqref{eq:LnPeSt}, the D-SCLF decoder updates the flip set in each additional decoding attempt, providing it with FHECC. 
However, this decoder still suffers from two notable issues.  
First, the newly proposed flip metric in Eq. \eqref{eq:LnPeSt} introduces significantly more exponential and logarithmic computations compared to the flip metric in Eq. \eqref{eq:Ei}.  
Second, the complexity of the decoder is highly sensitive to noise, particularly in low SNR environments, where its complexity becomes significantly higher than that of the CA-SCL decoder with comparable performance.

The recently proposed DL-D-SCLF decoder \cite{[181]} effectively simplifies the logarithmic and exponential computations in the flip metric of Eq. \eqref{eq:LnPeSt} by leveraging deep learning and Taylor series expansions.
Simulation results demonstrate that this simplification does not degrade performance.  
However, a major drawback is that the DL-D-SCLF decoder still fails to address the second issue mentioned above. 
Moreover, changes in the correction order or list size require the algorithm to retrain a new set of parameters through deep learning, which inevitably limits its flexibility.   

Addressing these challenges, developing a list flip decoder that achieves low average computational complexity while maintaining FHECC remains an open research problem.  
Such a decoder would represent a highly practical solution to meet the growing demands of high-reliability and low-energy communication systems.

\section{PC-DSCLF Decoder}
This section proposes the PC-DSCLF decoder, which primarily consists of two parts: a simplified flip metric and a hybrid-check-based decoding scheme.  
Notably, parts of this work have already been published in a preprint \cite{[182]}.

\subsection{Simplified Flip Metric for the D-SCLF Decoder}

Although the new metric of the DL-D-SCLF decoder effectively reduces the exponential and logarithmic computations in the flip metric of Eq. \eqref{eq:LnPeSt}, its parameters vary with changes in the list size and error-correction order.  
This variation creates obstacles for its application in diverse environments.  

To address this limitation, this subsection proposes another novel simplified flip metric based on Eq. \eqref{eq:LnPeSt}, considering the approximation in \cite{[22]} and the characteristics of list flip decoding algorithms.  
This flip metric satisfies 
\begin{equation}
{M^*}_{\beta}^{(i)}(\mathcal{S}_t) =E_{1}^{(i)}+\sum_{ k \leq i, k\in\mathcal{A}\setminus\mathcal{A}' }^{} f_{\beta=0.4}^*(E_{1}^{(k)})
+\sum_{k\in \mathcal{S}_t}^{}E_{1}^{(k)},
\label{eq:M*}
\end{equation}

where $z$ is a positive integer and 
\begin{equation}
f_{\beta=0.4}^*(x)= \left\{\begin{matrix} 1, & \text{if}\quad \left| x \right|  \leq z; \\
0, & \text{otherwise}. \\
\end{matrix}\right.
\end{equation}

To make \eqref{eq:M*} easier to understand, we replace ${M^*}_{\beta}^{(i)}(\mathcal{S}_t)$ with ${M}_{\beta}^*(\mathcal{S}_t\cup \{i\})$. Specifically, ${M}_{\beta}^*(\mathcal{S}_t)$ satisfies 
\begin{equation}
{M}_{\beta}^*(\mathcal{S}_t) =\sum_{k\in \mathcal{S}_t}E_{1}^{(k)}+\sum_{ k \leq i_t, k\in\mathcal{A}\setminus\mathcal{A}'} f_{\beta=0.4}^*(E_{1}^{(k)}),
\label{eq:M*simplify} 
\end{equation} 
where $i_t$ is the last element in $\mathcal{S}_t$.  

\begin{figure}[!t]
\centering
\includegraphics[scale=1]{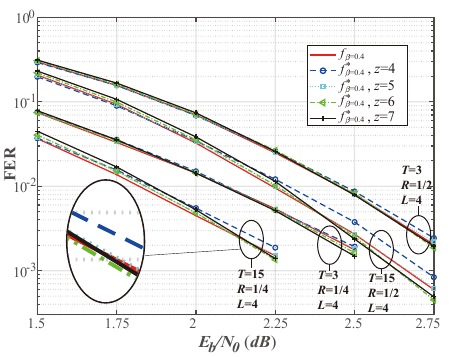} 
\caption{FER performance comparison of D-SCLF2 with different conditions. }
\label{fig2} 
\end{figure}

To explore the impact of different $z$ values on performance and to select a suitable $z$ value, we present Figure~\ref{fig2}, which shows a performance comparison of the D-SCLF2 decoder using the original $f_{\beta=0.4}$ metric and the improved $f_{\beta=0.4}^*$ metric with different $z$ values for $PC (512, 256+24)$.  
In Figure~\ref{fig2}, $L$, $T$, and $R$ refer to the list size, the number of additional decoding attempts, and the code rate, respectively.  
The simulation uses additive white Gaussian noise (AWGN) channels, BPSK modulation, and the Gaussian Approximation (GA) construction algorithm \cite{[25]} with a fixed $E_b/N_0$ of 4 dB.  
The generator polynomial of the 24 CRC bits is $g(x) = x^{24} + x^{23} + x^6 + x^5 + x + 1$.  
Note that D-SCLF2 refers to the D-SCLF decoder with the maximum number of flip indices for each additional decoding attempt set to 2.  
This setting facilitates comparison with the D-SCLF decoder in \cite{[20]} and helps evaluate the ability to correct high-order errors.  

As shown in Figure~\ref{fig2}, the frame error rate (FER) performance of the D-SCLF2 decoder using our flip metric is generally comparable to that of the original D-SCLF2 decoder.  
However, as $z$ increases, the FER performance of the D-SCLF2 decoder with $f_{\beta}^*(x)$ gradually deteriorates when $FER < 5\cdot 10^{-3}$.  
Considering the need for similar FER performance, it is preferable to use a fixed value for $z$.  
In the following sections, we set $z = 5$ as the default value.

\subsection{Hybrid Check Scheme for the D-SCLF Decoder}
The complexity of list flip decoders is closely related to the noise level. 
In high-noise scenarios, their complexity is often significantly higher than that of CA-SCL decoders with comparable FER performance.  
This phenomenon can be attributed to two main reasons.  
First, the probability of each information bit experiencing an event $\varepsilon$ increases, making it more difficult for the list flip decoding algorithm to locate the corresponding position within a limited number of extra attempts.  
Second, even if the list flip decoder successfully identifies the bit where the event $\varepsilon$ occurred and corrects the error, subsequent decoding steps may still be affected by noise, leading to new errors, which
also prevents the list flip decoding algorithm from finding the correct path within a finite number of additional decoding attempts.  
To address this issue, this subsection proposes a hybrid check scheme to narrow the error search range. 
This approach aims to improve the error detection accuracy in high-noise scenarios and enable early termination of decoding under such conditions.  

It is worth mentioning that a similar scheme has been validated in the PC-SC-Flip decoder \cite{[21]}.  
However, the proposed hybrid check scheme in this paper differs fundamentally from the one in \cite{[21]}.  
The PC-SC-Flip decoder focuses on enabling early termination, while its flip metric does not support FHECC. Specifically, high-order error-correction is only executed after all low-order corrections are completed, which inevitably limits the efficiency of flip operation.  
In contrast, the hybrid check scheme in this paper is proposed to integrate with the decoding process, enabling the decoding scheme to achieve FHECC while effectively reducing complexity.

To ensure the effectiveness of this hybrid check scheme, we establish the following requirements:
\begin{enumerate} 
	\item To ensure that multiple PC bits can effectively detect errors, all PC bits must belong to the predetermined set, and these PC bits should only protect bits contained within the predetermined set. 
	This is based on the findings in \cite{[23]}, which demonstrated that the predetermined set contains more than 99\% of all incorrect hard decisions caused by channel noise during the SC decoding process.
	\item To make PC bits play a greater role in early termination, the PC bits are dispersed among the non-frozen bits in the predetermined set, and the last PC bit should be placed at a certain distance from the CRC bits at the end of the non-frozen bit sequence.
\end{enumerate}

Based on the aforementioned requirements, the specific idea of the hybrid check scheme scheme is as follows:  
\begin{enumerate} 
	\item  Assume that  the vector $\mathbf{q}$ records the indices of the elements in the predetermined set within ${\mathbf{u}_\mathcal{A}}$, $n_{\mathbf{q}}$ is the number of elements in the predetermined set, the vector $\mathbf{p}$ records the indices of the PC bits in ${\mathbf{u}_\mathcal{A}}$, $n_{\mathbf{p}}$ is the number of PC bits, and $n_{\mathbf{c}}$ is the number of CRC bits. 
	\item Based on the second requirement, the sequence of non-frozen bits $\mathbf{u}_\mathcal{A}$ is divided into $n_{\mathbf{p}} + 1$ segments, ensuring that the last segment does not contain any PC bits. 
	\item Based on the first requirement, the positions of the non-frozen bits protected by PC bits are determined.  
\end{enumerate}
The specific implementation of this distribution scheme is described as follows.  
First, we define $n_1$ and $n_2$:
\begin{equation}
	\left\{\begin{matrix} n_1=\lceil n_{\mathbf{q}}/(n_{\mathbf{p}}+1)\rceil, \\
	n_2=\lfloor n_{\mathbf{q}}/(n_{\mathbf{p}}+1)\rfloor. \\
	\end{matrix}\right.
\end{equation}

If $n_1=n_2$,
\begin{equation}
	\mathbf{p}[i]=\mathbf{q}[i\cdot n_1],
	\end{equation}
and 
\begin{equation}
{\mathbf{u}_\mathcal{A}}[\mathbf{p}[i]]=\oplus_{k=(i-1)\cdot n_1+1}^{i\cdot n_1-1}{\mathbf{u}_\mathcal{A}}[\mathbf{q}[k]].
\end{equation}

If $n_1=n_2+1$,
\begin{equation}
	\mathbf{p}[i]= \left\{\begin{matrix} \mathbf{q}[i\cdot n_1], & \text{if}\quad i \leq c_1, \\
		\mathbf{q}[c_1\cdot n_1 + (i-c_1)\cdot n_2], & \text{if} \quad i>c_1, \\
		\label{eq:begin}
	\end{matrix}\right.
\end{equation}
where
\begin{equation}
	\left\{\begin{matrix}c_1=n_{\mathbf{q}}-n_2\cdot(n_{\mathbf{p}}+1), \\
		c_2=n_{\mathbf{p}}+1-c_1. \\
	\end{matrix}\right.
\end{equation}
Thus,
\begin{equation}
	{\mathbf{u}_\mathcal{A}}[\mathbf{p}[i]]= \left\{\begin{matrix} \oplus_{k=(i-1)\cdot n_1+1}^{i\cdot n_1-1}{\mathbf{u}_\mathcal{A}}[\mathbf{q}[k]], & \text{if}\quad i \leq c_1, \\
		\oplus_{k=k_1} ^{k_2}{\mathbf{u}_\mathcal{A}}[{\mathbf{q}[k]}], & \text{if} \quad i>c_1, \\
	\end{matrix}\right.
\end{equation}
where
\begin{equation}
	\left\{\begin{matrix}k_1=c_1\cdot n_1+(i-c_1-1)\cdot n_2+1, \\
		k_2=k_1+n_2-2. \\
	\end{matrix}\right.
	\label{eq:end}
\end{equation}
Note that $\oplus_{k=1}^{3}{\mathbf{u}_\mathcal{A}}[\mathbf{q}[k]]={\mathbf{u}_\mathcal{A}}[{\mathbf{q}[1]}] \oplus {\mathbf{u}_\mathcal{A}}[{\mathbf{q}[2]}] \oplus {\mathbf{u}_\mathcal{A}}[{\mathbf{q}[3]}]$.

\begin{figure}[!t]
\centering
\includegraphics[scale=1]{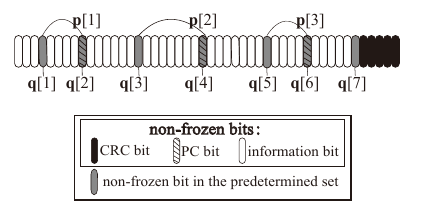} 
\caption{An example of our allocation schemes with 3 PC bits for a hypothetical non-frozen bits sequence.}
\label{fig3} 
\end{figure}

To explain our method more intuitively, we describe Figure~\ref{fig3} based on a hypothetical non-frozen bit sequence.  
In this figure, different blocks represent different non-frozen bits, with gray blocks indicating that the non-frozen bit belongs to the predetermined set.  
These non-frozen bits are divided into three categories: black blocks represent CRC bits, white blocks represent information bits, and the remaining blocks represent PC bits.  
Assume $n_{\mathbf{q}}=7$, $n_{\mathbf{p}}=3$, and $n_{\mathbf{c}}=5$.  
Through simple calculations, we obtain $(n_{\mathbf{q}},n_{\mathbf{p}},n_1,n_2,c_1,c_2 )=(7,3,2,1,3,1)$.  
By further calculations based on Eq. \eqref{eq:begin}-\eqref{eq:end}, we can obtain
$\mathbf{u}_\mathcal{A}[\mathbf{p}[1]]={\mathbf{u}_\mathcal{A}}[{\mathbf{q}[2]]}={\mathbf{u}_\mathcal{A}}[{\mathbf{q}[1]}]$, ${\mathbf{u}_\mathcal{A}}[\mathbf{p}[2]]={\mathbf{u}_\mathcal{A}}[{\mathbf{q}[4]}]={\mathbf{u}_\mathcal{A}}[{\mathbf{q}[3]}]$, and ${\mathbf{u}_\mathcal{A}}[\mathbf{p}[3]]={\mathbf{u}_\mathcal{A}}[{\mathbf{q}[6]}]= {\mathbf{u}_\mathcal{A}}[{\mathbf{q}[5]}]$.

\subsection{Details of the PC-DSCLF decoder}

This subsection provides the specific details of the proposed PC-DSCLF decoder.  
In addition to adopting the simplified flip metric introduced earlier, this decoder fully leverages the characteristics of the proposed hybrid check scheme to achieve a balance between error-correction capability and computational complexity.

\begin{algorithm}[!t]
	\caption{PC-DSCLF decoder}
	\label{alg:alg1}
	\KwIn{$ \mathcal{A}^c, K, L, T, N$}
	\KwOut{$\hat{u}_{1}^{N}$}
	$(c_c,c_p,\mathbf{e} _{1},\mathcal{S},\mathcal{M}) \leftarrow (1,0,\mathbf{0}_{1 \cdot N},\{0\}_{1 \cdot T},\mathbf{0}_{1 \cdot T}$); \\
	\For{$t\leftarrow$ 0 to $T$}{
	\For{$i\leftarrow$ 1 to $N$}{
	$\mathcal{L}_{best}^{(i)}\leftarrow$ \textbf{LFD}($\mathcal{S}_t$);\\
	\If{$i\in\mathcal{A}\setminus\mathcal{A}' $}{
	$\mathbf{e} _{1}[i] \leftarrow E_{1}^{(i)}$;
	}
	$c_p \leftarrow$ \textbf{PC}($\mathcal{L}_{best}^{(i)},c_p,i$);\\
	\If {$c_p=1$}{
	break;
	}
	$c_c \leftarrow$ \textbf{CRC}($\mathcal{L}_{best}^{(i)},c_c,i,N$);
	}
	\eIf{$c_c=1$}{
	$\hat{u}_1^N \leftarrow$ the path with smallest PM in $\mathcal{L}_{best}^{(N)}$;\\
	$(\mathcal{S},\mathcal{M}) \leftarrow$ \textbf{UPD}$(i,\mathcal{S},\mathcal{M},\mathbf{e}_{1},N,t,T)$;
	}
	{
	$\hat{u}_1^N \leftarrow$ the path with the smallest PM among those paths passing the CRC in $\mathcal{L}_{best}^{(N)}$;\\
	break;
	}
	}
	\Return{ $\hat{u}_1^N$ }
	\end{algorithm}
	
Algorithm 1 provides the detailed implementation of the PC-DSCLF decoder, where $T$ denotes the maximum number of additional attempts, $N$ represents the code length, and $\hat{u}_{1}^{N}$ is the estimated value of $u_{1}^{N}$. $c_c$ is a parameter reflecting the CRC result and is initialized to 1.
$c_p$ is a parameter reflecting the PC result and is initialized to 0.  
$\mathbf{e}_{1}$ is a vector used to store the $E_{1}$ values generated during the decoding process, initialized as a zero vector of length $N$, denoted by $\mathbf{0}_{1 \cdot N}$.  
$\mathcal{S}=\{\mathcal{S}_1,\mathcal{S}_2,\dots,\mathcal{S}_T\}$ refers to the list of flip sets, with each flip set initially set to $\{0\}$.  
We define the initial state of $\mathcal{S}$ as $\{0\}_{1 \cdot T}$.  
specially, $\mathcal{S}_0=\emptyset$.
$\mathcal{M}=\{M_\beta^*(\mathcal{S}_1),M_\beta^*(\mathcal{S}_2),\dots,M_\beta^*(\mathcal{S}_T)\}$ records the flip metrics of the flip sets in $\mathcal{S}$ and is initialized as $\mathbf{0}_{1 \cdot T}$.  
It is important to note that $\mathcal{M}$ must satisfy the following inequality:
\begin{equation}
M_\beta^*(\mathcal{S}_1) < M_\beta^*(\mathcal{S}_2) < \dots < M_\beta^*(\mathcal{S}_T),
\end{equation}
which ensures that flip sets with smaller flip metrics are prioritized for flipping.

\textbf{LFD}($\emptyset$) denotes a list flip decoder with a flip set $\emptyset$, whose decoding process is equivalent to that of the standard CA-SCL decoder.
\textbf{LFD}($\mathcal{S}_t$) denotes standard CA-SCL decoding during which the path selection at the bit indices given in this set $\mathcal{S}_t$ is flipped.
The flip operation refers to the current $\mathcal{L}_{best}^{(i)}$ is achieved by selecting $L$ paths with bigger PM from $\mathcal{L}^{(i)}$.
\textbf{PC}($\cdot$) denotes a PC function that verifies whether there exists a path in the current $\mathcal{L}_{best}$ that satisfies all PC checks up to the current decoding bit. If such a path exists, the function outputs $c_p=0$; otherwise, it outputs $c_p=1$.  
Similarly, \textbf{CRC}($\cdot$) denotes a CRC function that verifies whether there exists a path in $\mathcal{L}_{best}^{(N)}$ that passes the CRC check. If such a path exists, the function outputs $c_c=0$; otherwise, it outputs $c_c=1$.
\textbf{UPD}($\cdot$) denotes an update function that updates $\mathcal{S}$ and $\mathcal{M}$ based on the input conditions.

\begin{algorithm}[!t]
  \caption{\textbf{PC}}
  \label{alg:alg2}
  \KwIn{$\mathcal{L}_{best}^{(i)},c_p,i$}
  \KwOut{$c_p$}
  \If{the $i^{th}$ bit is a PC bit}{
  \eIf{all paths in $\mathcal{L}_{best}^{(i)}$ cannot pass all PCs before the $(i+1)^{th}$ bit}{
  $c_p \leftarrow 1$;
  }{
  $c_p \leftarrow 0$;
  }
  }
  \Return{ $c_p$ }
  \end{algorithm}
  
  \begin{algorithm}[!t]
  \caption{\textbf{CRC}}
  \label{alg:alg3}
  \KwIn{$\mathcal{L}_{best}^{(i)},c_c,i,N$}
  \KwOut{$c_c$}
  \If{$i=N$}{
  \eIf{all paths in $\mathcal{L}_{best}^{(N)}$ cannot pass the CRC}{
  $c_c \leftarrow 1$;
  }{
  $c_c \leftarrow 0$;
  }
  }
  \Return{ $c_c$ }
  \end{algorithm}

When $t=0$, the PC-DSCLF decoder executes \textbf{LFD}($\emptyset$) to obtain $\mathcal{L}_{best}$, and uses $\mathbf{e}_{1}$ to store the $E_{1}$ values corresponding to the bits in $\mathcal{A}\setminus\mathcal{A}'$.  
If the current decoding bit is a PC bit, the function \textbf{PC}($\cdot$) is activated.  
\begin{itemize}
    \item If the function \textbf{PC}($\cdot$) outputs $c_p=1$, the decoding attempt ends, and the function \textbf{UPD}($\cdot$) is activated to update $\mathcal{S}$ and $\mathcal{M}$ in preparation for the next decoding attempt. 
    \item If the function \textbf{PC}($\cdot$) continuously outputs $c_p=0$, the verification step of the function \textbf{CRC}($\cdot$) will be activated after decoding all the bits, and it will output an updated $c_c$ value. If $c_c=0$, the PC-DSCLF decoder outputs the path with the smallest PM among the paths passing the CRC in $\mathcal{L}_{best}^{(N)}$. Otherwise, the function \textbf{UPD}($\cdot$) is activated to update $\mathcal{S}$ and $\mathcal{M}$ for the next decoding attempt.  
\end{itemize}

When $t\neq 0$, the decoding steps of the PC-DSCLF decoder are generally similar to those for $t=0$. The main difference is that the PC-DSCLF decoder executes \textbf{LFD}($\mathcal{S}_t$) to obtain the corresponding parameters, and the process of updating $\mathcal{S}$ and $\mathcal{M}$ via the function \textbf{UPD}($\cdot$) differs. This difference is specifically presented in Algorithm 4.

Besides, we present and prove \emph{Theorem} 1 to ensure that the newly inserted flip set in our algorithm is executed only after the $t^{th}$ decoding attempt.
\begin{theorem}
	The PC-DSCLF decoder cannot execute the inserted flip set ${\mathcal{S}_t \cup {j} }$ before the $t^{th}$ additional attempt.
\end{theorem}
\textit{proof:}
	\begin{equation}
		\begin{aligned}
			&{M}_\beta^*(\mathcal{S}_t\cup \{j\}) - {M}_{\beta}^*(\mathcal{S}_t) \\
			&= E_{1}^{(j)}+\sum_{k\in \mathcal{S}_t}^{}E_{1}^{(k)}+\sum_{ k \leq j, k\in\mathcal{A}\setminus\mathcal{A}'  }^{} f_{\beta=0.4}^*(E_{1}^{(k)}) - \sum_{k\in \mathcal{S}_t}^{}E_{1}^{(k)} - \sum_{ k \leq i_t, k\in\mathcal{A}\setminus\mathcal{A}'  }^{} f_{\beta=0.4}^*(E_{1}^{(k)}) \\
			& = E_{1}^{(j)} + \sum_{i_t<k \leq j, k\in\mathcal{A}\setminus\mathcal{A}' } f_{\beta=0.4}^*(E_{1}^{(k)}).\nonumber
		\end{aligned}
	\end{equation}\ 
\noindent Since $E_{1}^{(j)}=ln\frac{\sum_{l=1}^{L}e^{-PM_l^{(j)}}}{\sum_{l=1}^{L}e^{-PM_{l+L}^{(j)}}} >0$ and $ f_{\beta=0.4}^*  \geq 0$, we can obtain ${M}_\beta^*(\mathcal{S}_t\cup \{j\}) > {M}_\beta^*(\mathcal{S}_t) $, which means the proof is completed.  \quad    $\hfill\blacksquare$

\begin{algorithm}[!t]
\caption{ \textbf{UPD}}
\label{alg:alg4}
\KwIn{$breakpoint,\mathcal{S},\mathcal{M},\mathbf{e}_{1},N,t,T$}
\KwOut{$(\mathcal{S},\mathcal{M})$}
$range$=min$(breakpoint,N)$;\\
\uIf {t=0}{
\For {$j \leftarrow $ $log_2(L)+1$ to $range$}{
\If {$j \in \{ \mathcal{A}\setminus\mathcal{A}' \} $}{
compute $M_\beta^*(\{j\})$ according to \eqref{eq:M*simplify};
}
}
$\mathcal{S} \leftarrow $ $T$ indexes of non-frozen bits with smaller $M_\beta^*$, and these non-frozen bits is in $\{ \mathcal{A}\setminus\mathcal{A}' \} $. \\
$\mathcal{M} \leftarrow $ $\{ M_\beta^*(\mathcal{S}_1),M_\beta^*(\mathcal{S}_2),...,M_\beta^*(\mathcal{S}_T) \}$ \\ 
\footnotesize //$M_\beta^*(\mathcal{S}_1) <M_\beta^*(\mathcal{S}_2)<...<M_\beta^*(\mathcal{S}_T)$ \normalsize 
}

\ElseIf{$0<t<T$}{
$i_t \leftarrow $ the last element in $\mathcal{S}_t$;\\
\For{$j \leftarrow $ $i_t+1$ to $range$}{
\If{$j \in \{ \mathcal{A}\setminus\mathcal{A}' \} $}{
\If{${M}_\beta^*(\mathcal{S}_t\cup \{j\}) < {M}_\beta^*(\mathcal{S}_T)$}{
$\mathcal{S} \leftarrow $ $\{ \mathcal{S}_1,...,\mathcal{S}_t,...,\mathcal{S}_t\cup \{j \},..., \mathcal{S}_{T-1} \}$;\\
\footnotesize // the new $\mathcal{S}_T$ in the current $\mathcal{S}$ is the $\mathcal{S}_{T-1}$ in the $\mathcal{S}$ before being updated \normalsize \\
$\mathcal{M} \leftarrow $ $\{ M_\beta^*(\mathcal{S}_1),...,M_\beta^*(\mathcal{S}_t),...,M_\beta^*(\mathcal{S}_t \cup \{j \}),...,M_\beta^*(\mathcal{S}_{T-1}) \}$;\\
\footnotesize //$M_\beta^*(\mathcal{S}_1)<...<M_\beta^*(\mathcal{S}_t)<...<M_\beta^*(\mathcal{S}_t \cup \{j \})<...<M_\beta^*(\mathcal{S}_{T-1}) $ \normalsize
}
}
} 
}
\Return{ $(\mathcal{S},\mathcal{M})$ }
\end{algorithm}

\section{Simulation Results and Discussions}

This section compares the proposed PC-DSCLF algorithm with the latest algorithms that feature FHECC under different conditions.
In these performance comparisons, BPSK modulation is used to transmit codewords over AWGN channels, and the GA algorithm with 4 dB is employed to construct polar codes.  
Without loss of generality, we adopt a check bit structure and consistent with that in \cite{[21]}.  
The total number of check bits, $n_C$, satisfies $n_C=n_c+n_p=24$, and there are two cases:  
\begin{itemize}
    \item Case 1: 8 PC bits and 16 CRC bits. 
    \item Case 2: 24 CRC bits. 
\end{itemize}
The corresponding generator polynomials of Case 1 and Case 2 are $ g(x)=x^{16}+x^{15}+x^2+1$ and $g(x)=x^{24}+x^{23}+x^6+x^5+x+1$, respectively.
In addition, we used the average cumulative number of paths over $\mathcal{A}$ to measure computational complexity, as the proposed PC-DSCLF decoder terminates decoding early when an error is detected 
For simplicity, we denote this parameter as $D$, which satisfies the following equation:
\begin{equation}
	D=\frac{
		\sum_{i=1}^{n_F} ( \sum_{t=0}^{T} ( \left( n_{t,i} - \log_2 L_{t,i} \right) L_{t,i} ) + \sum_{j=1}^{L_{t,i}} 2^j )
		}{
		n_F
		},
\end{equation}
where $n_F$ represents the total number of frames decoded by the decoder, $T$ is the maximum number of additional decoding attempts, $n_{t,i}$ denotes the number of non-frozen bits decoded during the $t^{\text{th}}$ additional decoding attempt for the $i^{\text{th}}$ frame, and $L_{t,i}$ represents the width of the candidate path list during the $t^{\text{th}}$ additional decoding attempt for the $i^{\text{th}}$ frame.
Specifically, $t=0$ indicates that the decoder is performing the initial decoding attempt.
Since $\log_2 L_{t,i}$ is much smaller than $n_{t,i}$ in practice, and $L_{t,i}$ is fixed to $L$ in this case, $D$ can be further simplified as follows:
\begin{equation}
	D=L \cdot \frac{
		\sum_{i=1}^{n_F}  \sum_{t=0}^{T} n_{t,i}  
		}{
		n_F
		}.
\end{equation}
To illustrate $D$, we provide a simple example: for $PC(512, 256+24)$, the value of $D$ in a standard CA-SCL decoder ($L=4$) is approximately $4 \cdot (256+24)$.

\begin{figure}[!t]
  \centering
  \includegraphics[scale=0.8]{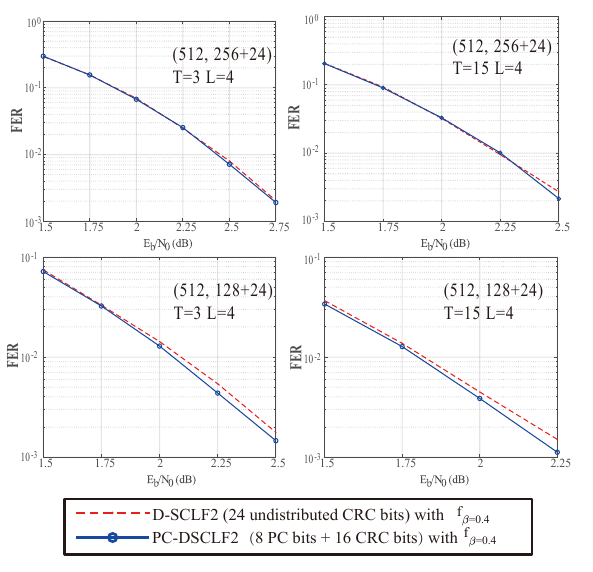} 
  \caption{FER performance comparison between D-SCLF2 (24 undistributed CRC bits) and PC-DSCLF2 (8 PC bits and 16 undistributed CRC bits), both with the same $f_{\beta=0.4}$ and $L=4$.}
  \label{fig4} 
  \end{figure}
  
  \begin{figure}[!t]
    \centering
    \includegraphics[scale=0.8]{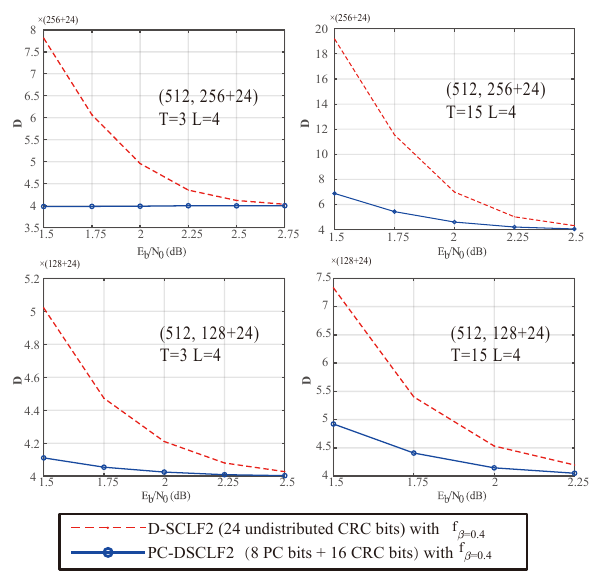} 
    \caption{ $D$ comparison between D-SCLF2 (24 undistributed CRC bits) and PC-DSCLF2 (8 PC bits and 16 undistributed CRC bits), both with the same $f_{\beta=0.4}$ and $L=4$. }
    \label{fig5} 
  \end{figure}

To demonstrate the effectiveness of our hybrid check scheme, we compare the performance of D-SCLF2 and PC-DSCLF2 in Figure~\ref{fig4} and Figure~\ref{fig5}, using the same $f_{\beta=0.4}$.  
Figure~\ref{fig4} presents the FER performance of D-SCLF2 (24 undistributed CRC bits) and PC-DSCLF2 (8 PC bits and 16 undistributed CRC bits), under the same $f_{\beta=0.4}$ and $L=4$.  
Under the same $R$, $L$, $T$, and $n_C$, the performance curves of the two algorithms are almost identical.  
Notably, at high $E_b/N_0$, the FER performance of PC-DSCLF2 is slightly better than that of D-SCLF2.  
This is because a larger number of CRC bits does not necessarily lead to better error-correction performance \cite{[24]}.  
Figure~\ref{fig5} shows the $D$ values of the algorithms presented in Figure~\ref{fig4}.  
Simulation results in Figure~\ref{fig5} indicate that ourhybrid check scheme with $f_{\beta=0.4}$ achieves up to $\frac{19.2-6.9}{19.2}=64.1\%$ average complexity reduction compared to the D-SCLF2 algorithm with 24 undistributed CRC bits, without any degradation in error-correction performance.  
Thus, the proposed hybrid check scheme maintains the error-correction performance of the D-SCLF2 algorithm while significantly reducing its computational complexity.

\begin{figure}[!t]
  \centering
  \includegraphics[scale=0.8]{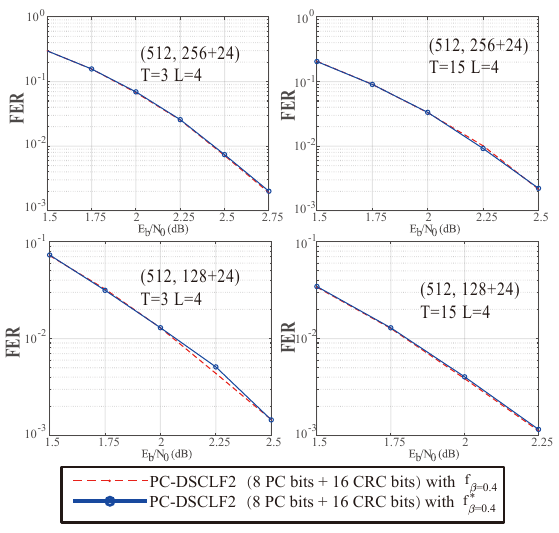} 
  \caption{ FER performance comparison of the PC-DSCLF2 (8 PC bits and 16 undistributed CRC bits) algorithms with different flip metrics. }
  \label{fig6} 
  \end{figure}

\begin{figure}[!t]
  \centering
  \includegraphics[scale=0.8]{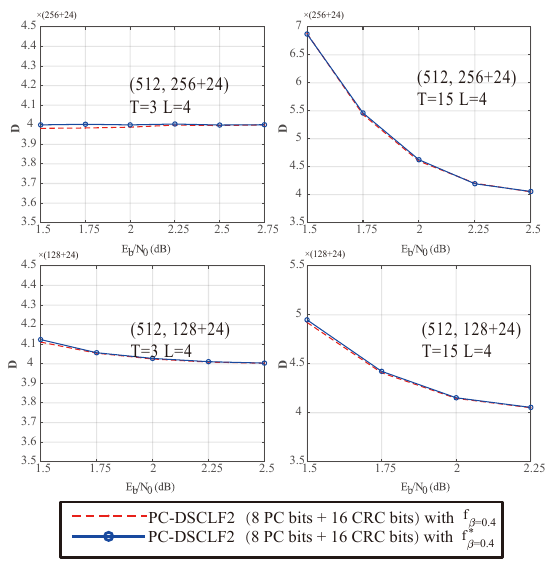} 
  \caption{ $D$ comparison of the PC-DSCLF2 (8 PC bits and 16 undistributed CRC bits) algorithms with different flip metrics. }
  \label{fig7} 
  \end{figure}

To demonstrate the effectiveness of our simplified flip metric, we present Figure~\ref{fig6} and Figure~\ref{fig7}, which show the FER performance and $D$ of the PC-DSCLF2 algorithm with different flip metrics, respectively.  
The performance curves of the PC-DSCLF2 decoder with the simplified flip metric and those with the original flip metric are nearly identical.  
Thus, our simplified flip metric eliminates the logarithmic and exponential operations in $f_{\beta=0.4}$, without increasing $D$ or degrading FER performance.

\begin{figure}[!t]
  \centering
  \includegraphics[scale=0.8]{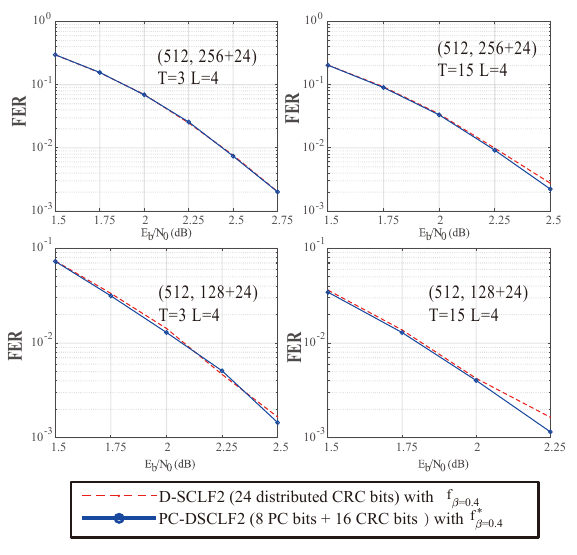} 
  \caption{ FER performance between D-SCLF2 (24 distributed CRC bits and $f_{\beta=0.4}$) and PC-DSCLF2 (8 PC bits and 16 undistributed CRC bits and $f_{\beta=0.4}^*$), both with $L=4$.}
  \label{fig8} 
  \end{figure}

\begin{figure}[!t]
  \centering
  \includegraphics[scale=0.8]{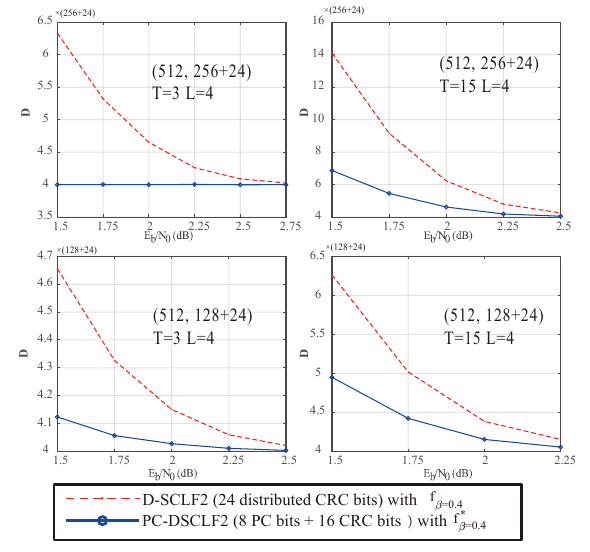} 
  \caption{ $D$ comparison between D-SCLF2 (24 distributed CRC bits and $f_{\beta=0.4}$) and PC-DSCLF2 (8 PC bits and 16 undistributed CRC bits and $f_{\beta=0.4}^*$), both with $L=4$.}
  \label{fig9} 
  \end{figure}

  To demonstrate the effectiveness of the PC-DSCLF decoder, we present Figure~\ref{fig8} and Figure~\ref{fig9}, which show the FER performance and $D$ of the D-SCLF2 (24 distributed CRC bits and $f_{\beta=0.4}$) and PC-DSCLF2 (8 PC bits, 16 undistributed CRC bits, and $f_{\beta=0.4}^*$), respectively.  
  For the sake of illustration, we define $D_{SCL}$ as the $D$ of a standard SCL decoder.  
  As $E_b/N_0$ increases, $D$ converges to $D_{SCL}$ regardless of the code rate or decoder.  
  In all cases, our PC-DSCLF decoder converges to $D_{SCL}$ faster than the D-SCLF decoder.  
  Specifically, our PC-DSCLF decoder ($T=15$ and $L=4$) achieves up to a $\frac{14.1-6.9}{14.1}=51.1\%$ reduction in $D$ compared to the D-SCLF decoder ($T=15$ and $L=4$) for PC(512, 256+24).  

\begin{figure}[!t]
	\centering
	\includegraphics[scale=0.8]{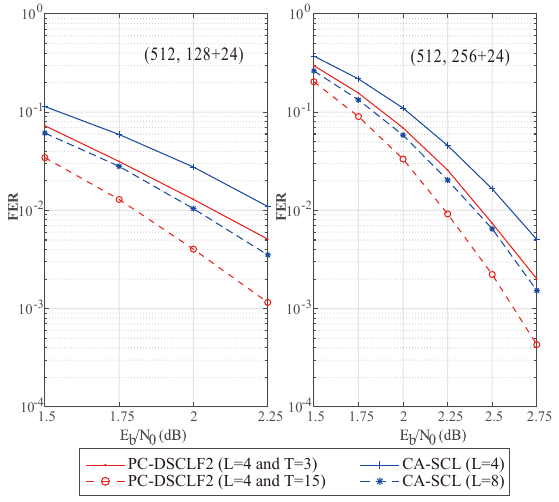} 
	\caption{ FER performance comparison between CA-SCL (24 distributed CRC bits) and PC-DSCLF2 (8 PC bits and 16 undistributed CRC bits). }
	\label{fig10} 
	\end{figure}

\begin{figure}[!t]
    \centering
    \includegraphics[scale=0.8]{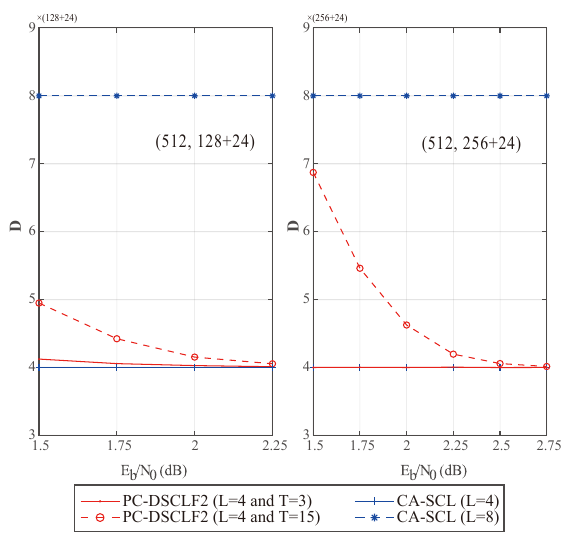} 
    \caption{ $D$ comparison between CA-SCL (24 distributed CRC bits) and PC-DSCLF2 (8 PC bits and 16 undistributed CRC bits).}
    \label{fig11} 
    \end{figure}

  To further illustrate the performance advantages of the PC-DSCLF decoder, we present Figure~\ref{fig10} and Figure~\ref{fig11}, which show the FER performance and $D$ of the CA-SCL decoder (24 distributed CRC bits) and the PC-DSCLF2 decoder (8 PC bits and 16 undistributed CRC bits), respectively.  
  With only three additional decoding attempts, the PC-DSCLF2 decoder achieves FER performance comparable to that of the CA-SCL decoder with $L=8$, while maintaining a similar $D$ to that of the CA-SCL decoder with $L=4$, for $PC(512, 256+24)$.  
  Furthermore, the PC-DSCLF decoder with $T=15$ demonstrates significantly lower computational complexity and better FER performance compared to the CA-SCL decoder with $L=8$.  
  Thus, the PC-DSCLF decoder, with its low computational complexity and reduced storage requirements, achieves the error-correction performance of the CA-SCL decoder, which has higher computational complexity and larger storage requirements.

In summary, compared with the latest algorithms that feature FHECC under different conditions, the proposed PD-DSCLF decoder achieves a better balance between error-correction capability and computational complexity.

\section{Conclusions}
Considering the future requirements for high-reliability and low-power communication, and the performance advantages of polar codes in short code lengths, this paper proposes a PC-DSCLF decoder with low complexity and high error-correction performance. 
Simulation results show that, compared to the latest list flip decoder with FHECC, the proposed PC-DSCLF algorithm significantly reduces complexity without compromising error-correction performance, particularly under low SNR conditions.

\section*{Acknowledgements}
This work is supported by “the Fundamental Research Funds for the Central Universities”.

\bibliographystyle{elsarticle-num} 
\bibliography{reference}






\end{document}